\documentclass[aps,prl,reprint,showpacs,superscriptaddress,bibnotes,twoside]{revtex4-1}

\usepackage{amsmath}
\usepackage{amssymb}
\usepackage{graphicx}
\usepackage{subfigure}
\usepackage{color}

\begin{document}

%\preprint{Preprint: PTMatQuaStack-PRL-Submission}

%%%%%%%%%%%%%%%%%%%%%%%%%%%%%%%%%%%%%%%%%%%%%%%%%%%%%%%%%%%%%%%%%%%%%
\title{$\mathcal{P}\mathcal{T}$-symmetric transport in non-$\mathcal{P}\mathcal{T}$-symmetric bi-layer optical arrays}
%%%%%%%%%%%%%%%%%%%%%%%%%%%%%%%%%%%%%%%%%%%%%%%%%%%%%%%%%%%%%%%%%%%%%

%%%%%%%%%%%%%%%%%%%%%%%%%%%%%%%%%%%%%%%%%%%%%%%%%%%%%%%%%%%%%%%%%%%%%
\author{J.~Ramirez-Hernandez}
\affiliation{Instituto de F\'{\i}sica, Benem\'erita Universidad Aut\'{o}noma de Puebla, Apdo. Post. J-48, Puebla, Pue., 72570, M\'{e}xico}
%%%%%%%%%%%%%%%%%%%%%%%%%%%%%%%%%%%%%%%%%%%%%%%%%%%%%%%%%%%%%%%%%%%%%

%%%%%%%%%%%%%%%%%%%%%%%%%%%%%%%%%%%%%%%%%%%%%%%%%%%%%%%%%%%%%%%%%%%%%
\author{F.~M.~Izrailev}
\email{felix.izrailev@gmail.com}
\affiliation{Instituto de F\'{\i}sica, Benem\'erita Universidad Aut\'{o}noma de Puebla, Apdo. Post. J-48, Puebla, Pue., 72570, M\'{e}xico}
%%%%%%%%%%%%%%%%%%%%%%%%%%%%%%%%%%%%%%%%%%%%%%%%%%%%%%%%%%%%%%%%%%%%%

%%%%%%%%%%%%%%%%%%%%%%%%%%%%%%%%%%%%%%%%%%%%%%%%%%%%%%%%%%%%%%%%%%%%%
\author{N.~M.~Makarov}
\email{makarov.n@gmail.com}
\affiliation{Instituto de Ciencias, Benem\'erita Universidad Aut\'{o}noma de Puebla, \\Priv. 17 Norte No. 3417, Col. San Miguel Hueyotlipan, Puebla, Pue., 72050, M\'{e}xico}
%%%%%%%%%%%%%%%%%%%%%%%%%%%%%%%%%%%%%%%%%%%%%%%%%%%%%%%%%%%%%%%%%%%%%

%%%%%%%%%%%%%%%%%%%%%%%%%%%%%%%%%%%%%%%%%%%%%%%%%%%%%%%%%%%%%%%%%%%%%
\author{D.~N.~Christodoulides}
\affiliation{College of Optics and Photonics-CREOL, University of Central Florida, 32816, USA}
%%%%%%%%%%%%%%%%%%%%%%%%%%%%%%%%%%%%%%%%%%%%%%%%%%%%%%%%%%%%%%%%%%%%%

\date{\today}

%%%%%%%%%%%%%%%%%%%%%%%%%%%%%%%%%%%%%%%%%%%%%%%%%%%%%%%%%%%%%%%%%%%%%2
\begin{abstract}
We study transport properties of an array created by alternating $(a,b)$ layers with balanced loss/gain characterized by the key parameter $\gamma$. It is shown that for non-equal widths of $(a,b)$ layers, i.e., when the corresponding Hamiltonian is non-$\mathcal{P}\mathcal{T}$-symmetric, the system exhibits the scattering properties similar to those of truly $\mathcal{P}\mathcal{T}$-symmetric models provided that without loss/gain the structure presents the matched quarter stack. The inclusion of the loss/gain terms leads to an emergence of a finite number of spectral bands characterized by real values of the Bloch index. Each spectral band consists of a central region where the transmission coefficient $T_N\geqslant1$, and two side regions with $T_N\leqslant1$. At the borders between these regions the unidirectional reflectivity occurs. Also, the set of Fabry-Perrot resonances with $T_N=1$ are found in spite of the presence of loss/gain.
\end{abstract}
%%%%%%%%%%%%%%%%%%%%%%%%%%%%%%%%%%%%%%%%%%%%%%%%%%%%%%%%%%%%%%%%%%%%%

%%%%%%%%%%%%%%%%%%%%%%%%%%%%%%%%%%%%%%%%%%%%%%%%%%%%%%%%%%%%%%%%%%%%%
%11.30.Er Charge conjugation, parity, time reversal, and other discrete symmetries
%42.25.Bs Wave propagation, transmission and absorption
%42.82.Et Waveguides, couplers, and arrays
%%%%%%%%%%%%%%%%%%%%%%%%%%%%%%%%%%%%%%%%%%%%%%%%%%%%%%%%%%%%%%%%%%%%%

\maketitle

%%%%%%%%%%%%%%%%%%%%%%%%%%%%%%%%%%%%%%%%%%%%%%%%%%%%%%%%%%%%%%%%%%%%%
\textbf{Introduction -}
%%%%%%%%%%%%%%%%%%%%%%%%%%%%%%%%%%%%%%%%%%%%%%%%%%%%%%%%%%%%%%%%%%%%%
The possibility of deliberately intermixing loss and gain in optical structures as a means to attain physical properties that are otherwise out of reach in standard arrangements has recently been explored in a number of studies \cite{1,2,3,4,5,6,L10,8}. In general, such non-conservative configurations can display surprising behavior resulting from the possible presence of exceptional points \cite{11,12} and mode bi-orthogonality. A particular class of non-Hermitian systems that has recently attracted considerable attention is that respecting parity-time ($\mathcal{P}\mathcal{T}$) symmetry \cite{1,2,3,4,5}. In this regard, if a Hamiltonian is $\mathcal{P}\mathcal{T}$-symmetric, then its ensuing spectrum happens to be entirely real in spite of the fact that the potential involved is complex \cite{11,12}. Once however a loss/gain parameter exceeds a critical value, this symmetry can spontaneously break and consequently the eigenvalue spectrum ceases to be real and hence enters the complex domain \cite{11}.

The availability of optical attenuation and amplification makes optics an ideal ground where $\mathcal{P}\mathcal{T}$-symmetric effects can be experimentally observed and investigated \cite{3,4,13}. In the paraxial regime, a necessary (but not sufficient) condition to impose this symmetry in photonics is that the complex refractive index distribution obeys $n(x)=n^*(-x)$ \cite{1,2}. This latter condition directly implies real part of the refractive index potential must be an even function of position while its imaginary component must be antisymmetric.

Thus far, the ramifications of parity-time symmetry on optical transport have been systematically considered in several theoretical and experimental investigations. Processes emerging from PT-symmetry, like double refraction, band merging, unidirectional invisibility, abrupt phase transitions and nonreciprocal wave propagation have been predicted and observed \cite{1,3,13,Lo11,15,16}. In addition this same symmetry was recently utilized to enforce single-mode behavior in laser micro-cavities as well as to judiciously control wave dynamics around exceptional points and photonic crystals \cite{17}. Finally, similar concepts are nowadays used in other fields, like for example opto-mechanics, acoustics, nonlinear optics, plasmonics, metamaterials, and imaging to mention a few \cite{23}. In view of these developments, the following question naturally arises. To which extent could a system depart from an exact parity-time-symmetry and still exhibit $\mathcal{P}\mathcal{T}$-symmetric characteristics? Recently, a possibility of synthesizing complex potentials that support entirely real spectra despite the fact that they violate the necessary condition for $\mathcal{P}\mathcal{T}$-symmetry has been discussed in Ref.~\cite{MHC13} within the context of optical supersymmetry.

In this paper we show that it is indeed possible to observe $\mathcal{P}\mathcal{T}$-scattering behavior even in systems that strictly speaking lack this symmetry. This is explicitly demonstrated in perfectly matched multilayer non-Hermitian arrangements having unequal layer widths. In this case, the incorporation of loss/gain domains leads to a finite number of bands-all associated with a real Bloch index that is reminiscent of actual $\mathcal{P}\mathcal{T}$-symmetric lattices \cite{Lo11}. Special attention has been paid to an emergence of unidirectional reflectivity. Our results indicate that a $\mathcal{P}\mathcal{T}$-symmetric response can be quite robust and therefore can be observed in more involved structures even in the absence of a strict $\mathcal{P}\mathcal{T}$-symmetry.

\begin{figure}[h!!!]
\centering
\includegraphics[scale=0.25]{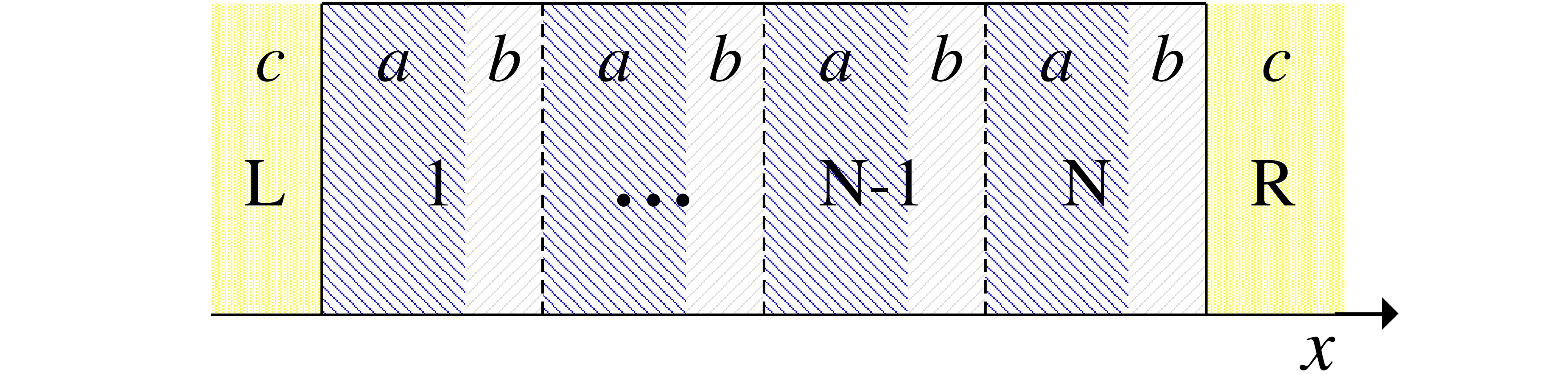}
\caption{(Color online) A sketch of the model.}\label{fig:Array-BQS}
\end{figure}

%%%%%%%%%%%%%%%%%%%%%%%%%%%%%%%%%%%%%%%%%%%%%%%%%%%%%%%%%%%%%%%%%%%%%
\textbf{The model -}
%%%%%%%%%%%%%%%%%%%%%%%%%%%%%%%%%%%%%%%%%%%%%%%%%%%%%%%%%%%%%%%%%%%%%
We consider the propagation of an electromagnetic wave of frequency $\omega$ through a regular array of $N$ identical unit $(a,b)$ cells embedded in a homogeneous $c\,$-medium, see Fig.~\ref{fig:Array-BQS}. Each cell consists of two dielectric $a$ and $b$ layers with the thicknesses $d_a$ and $d_b$, respectively, so that $d=d_a+d_b$ is the unit-cell size. The $a$ and $b$ layers are made of the materials absorbing and amplifying the electromagnetic waves, respectively.
%
%%%%%%%%%%%%%%%%%%%%%%%%%%%%%%%%%%%%%%%%%%%%%%%%%%%%%%%%%%%%%%%%%%%%%
%%%%%%%%%%%%%%%%%%%%%%%%%%%%%%%%%%%%%%%%%%%%%%%%%%%%%%%%%%%%%%%%%%%%%
%
The loss and gain in the layers are incorporated via complex permittivities, while the magnetic permeabilities $\mu_{a,b}$ are assumed to be real and positive constants. The refractive indices $n_{a,b}$, impedances $Z_{a,b}$ and wave phase shifts $\varphi_{a,b}$ of $a$ and $b$ layers are presented as follows,
\begin{eqnarray*}
n_a=n_a^{(0)}(1+i\gamma),\, Z_a=Z(1+i\gamma)^{-1},\,\varphi_a=\varphi(1+i\gamma)/2,\\
n_b=n_b^{(0)}(1-i\gamma),\, Z_b=Z(1-i\gamma)^{-1},\,\varphi_b=\varphi(1-i\gamma)/2,
\end{eqnarray*}
where
\begin{equation*}
Z=\mu_a/n_a^{(0)}=\mu_b/n_b^{(0)},\quad\varphi=2\omega n_a^{(0)}d_a/c=2\omega n_b^{(0)}d_b/c\,.
\end{equation*}
Here the parameter $\gamma>0$ measures the strength of balanced loss and gain inside $a$ and $b$ layers, respectively, and $\varphi=\varphi_{a}+\varphi_{b}$ is the \emph{purely real} phase shift of the wave passing each of the unit cells. In the case of no loss/gain ($\gamma=0$) the structure is known as the \emph{matched quarter stack} for which the $a$ and $b$ layers are perfectly matched and have equal optic paths, $n_a^{(0)}d_a=n_b^{(0)}d_b$. As one can see, for $\gamma \neq 0$ and $d_a \neq d_b$ the corresponding Hamiltonian has no the $\mathcal{P}\mathcal{T}$-symmetry. A particular $\mathcal{P}\mathcal{T}$-symmetric realization is achieved only when $d_a=d_b$ with $\varepsilon_a^{(0)}=\varepsilon_b^{(0)}$, and $\mu_a=\mu_b$.

Within any layer an electromagnetic wave propagates along the $x\,$-direction (Fig.~\ref{fig:Array-BQS}) perpendicular to the stratification according to the 1D Helmholtz equation. Its general solution inside every unit cell can be presented as a superposition of traveling forward and backward plane waves. By combining these solutions with the boundary conditions for the wave at the interfaces between $a$ and $b$ layers, one can obtain the \emph{unit-cell transfer matrix} $\hat{M}$ relating the wave amplitudes of two adjacent unit cells (see, e.g., Ref.~\cite{IKM12}). For our model the elements of the $M$-matrix read
\begin{subequations}\label{eq:PTMQS-M}
\begin{eqnarray}
&&M_{11}=\frac{\exp(i\varphi)+\gamma^2\exp(-\gamma\varphi)}{1+\gamma^2},\\
&&M_{12}=\frac{i\gamma}{1+\gamma^2}[\exp(-i\varphi)-\exp(\gamma\varphi)],\\
&&M_{21}=\frac{i\gamma}{1+\gamma^2}[\exp(i\varphi)-\exp(-\gamma\varphi)],\\
&&M_{22}=\frac{\exp(-i\varphi)+\gamma^2\exp(\gamma\varphi)}{1+\gamma^2}.
\end{eqnarray}
\end{subequations}
Note that $\det\hat{M}=1$ in accordance with the general condition inherent for transfer matrices. On the other hand, the transfer matrix $\hat{M}(\gamma)$ has specific symmetry,
\begin{equation}\label{eq:PTMQS-MTR}
M_{22}(\gamma)=M_{11}^*(-\gamma)\,,\qquad M_{21}(\gamma)=M_{12}^*(-\gamma)\,,
\end{equation}
where ``$*$" stands for the complex conjugation. This symmetry can be compared with that emerging for various $\mathcal{P}\mathcal{T}$-symmetric Hamiltonians \cite{L10,M13,VHIC14,VIC15}. Clearly, it differs from the standard one, $M_{22}=M_{11}^*$, $M_{21}=M_{12}^*$, that originates from time-reversal symmetry.

%%%%%%%%%%%%%%%%%%%%%%%%%%%%%%%%%%%%%%%%%%%%%%%%%%%%%%%%%%%%%%%%%%%%%
\textbf{Frequency band structure -}
%%%%%%%%%%%%%%%%%%%%%%%%%%%%%%%%%%%%%%%%%%%%%%%%%%%%%%%%%%%%%%%%%%%%%
The dispersion relation for the Bloch index $\mu_B$ determined through the eigenvalues $\exp(\pm i\mu_B)$ of the transfer $\hat{M}$-matrix, is defined by the matrix trace, $2\cos\mu_B=M_{11}+M_{22}$. Thus, one obtains
\begin{equation}\label{eq:PTMQS-DispEq}
\cos\mu_B=\frac{\cos\varphi+\gamma^2 \cosh(\gamma \varphi)}{1+\gamma^2}\,.
\end{equation}
This relation determines the dependence of the Bloch wave number $\kappa=\mu_B/d$ on the frequency $\omega$ (or, on the phase shift $\varphi\propto\omega$) for the wave inside the scattering region. One of the important conclusions that follows from Eq.~(\ref{eq:PTMQS-DispEq}), is a \emph{finite} number $N_{band}$ of \emph{spectral bands} defined by the condition $|\cos\mu_B|<1$. This number $N_{band}$ depends on the parameter $\gamma$ only, and for $\gamma\ll1$ can be estimated as
\begin{equation}
N_{band}\approx(1/\pi\gamma)\ln(1/\gamma) .
\end{equation}
Within these bands the Bloch index $\mu_B$ is real in spite of the presence of loss and gain. This property has been widely discussed in view of the nature of the $\mathcal{P}\mathcal{T}$-symmetric systems. As is known, the real-valued solutions $\mu_B$ to the dispersion equation $2\cos\mu_B=M_{11}+M_{22}$ can exist only when the trace $M_{11}+M_{22}$ of the transfer $\hat{M}$-matrix is real-valued, i.e. if
\begin{equation}\label{eq:PTMQS-ImM}
\mathrm{Im}M_{22}=-\mathrm{Im}M_{11}.
\end{equation}
In our model this determinative condition is provided not due to the $\mathcal{P}\mathcal{T}$-symmetry but merely due to the balance between loss and gain.

Outside the bands, where $|\cos\mu_B|>1$, the index $\mu_B$ is purely imaginary, thus creating  {\it spectral gaps}. Here the waves are known as the evanescent Bloch states, attenuated on the scale of the order of $|\mu_B|^{-1}$. Therefore, for a sufficiently long structure, $N|\mu_B|\gg1$, the transmission is exponentially small. It should be noted that in the absence of loss/gain ($\gamma=0$), there are no spectral gaps since all spectral bands are touching.
%
%%%%%%%%%%%%%%%%%%%%%%%%%%%%%%%%%%%%%%%%%%%%%%%%%%%%%%%%%%%%%%%%%%%%%
\begin{figure}[h!!!]
\centering
\includegraphics[scale=0.2]{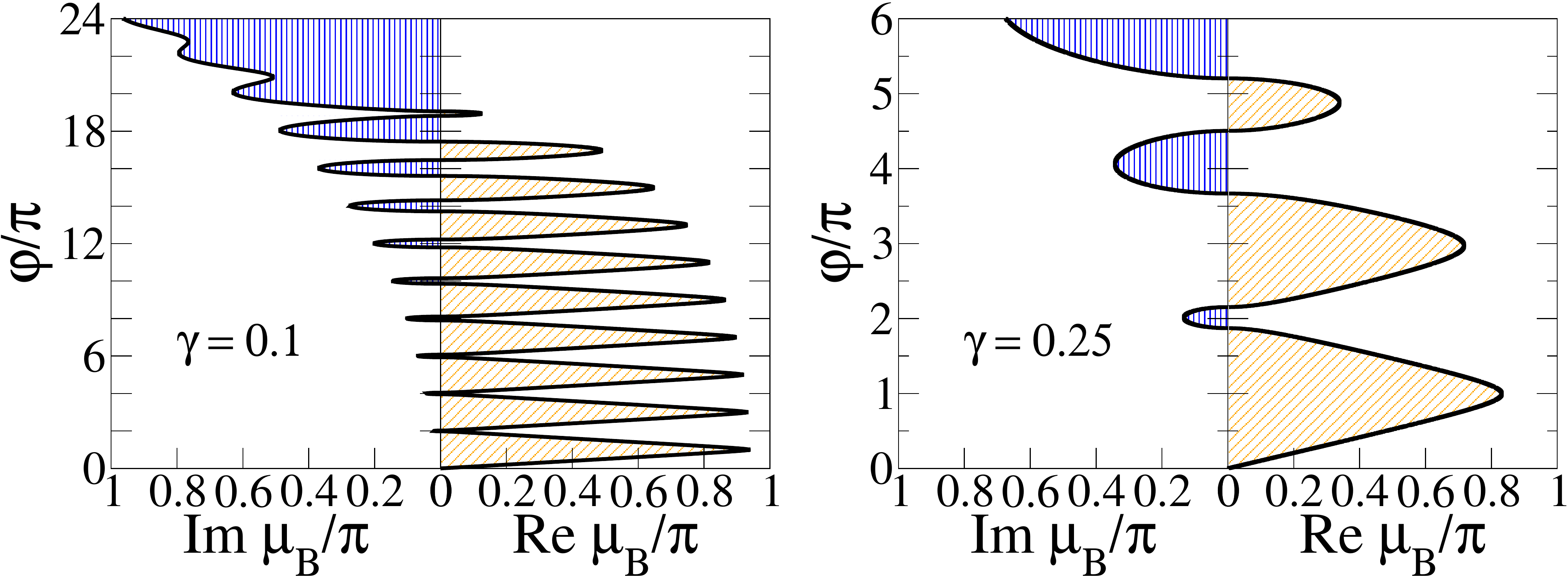}
\caption{The dependence \eqref{eq:PTMQS-DispEq} of the Bloch index $\mu_B$ on the phase shift $\varphi\propto\omega$ for two values of $\gamma$. Spectral bands and gaps are shown by shaded areas.}\label{fig:BandGap}
\end{figure}
%%%%%%%%%%%%%%%%%%%%%%%%%%%%%%%%%%%%%%%%%%%%%%%%%%%%%%%%%%%%%%%%%%%%%

When $\gamma$ exceeds the critical value $\gamma_{cr}=1$, there are no spectral bands since the Bloch index $\mu_B(\omega)$ becomes imaginary for any frequency $\omega$. Two examples of the dependence $\mu_B(\omega)$ are shown in Fig.~\ref{fig:BandGap} for different values of $\gamma$. It is quite instructive that within any spectral band the maximal value of the  Bloch index $\mu_B^{(max)}<\pi$ for any nonzero $\gamma$. The smaller the parameter $\gamma$, the closer to $\pi$  can be the Bloch index $\mu_B$. On the other hand, for any given $\gamma$ from the interval $0<\gamma<1$ the higher the band the smaller the value of $\mu_B^{(max)}$.

\begin{figure}[h!!!]
\includegraphics[width=0.8\columnwidth]{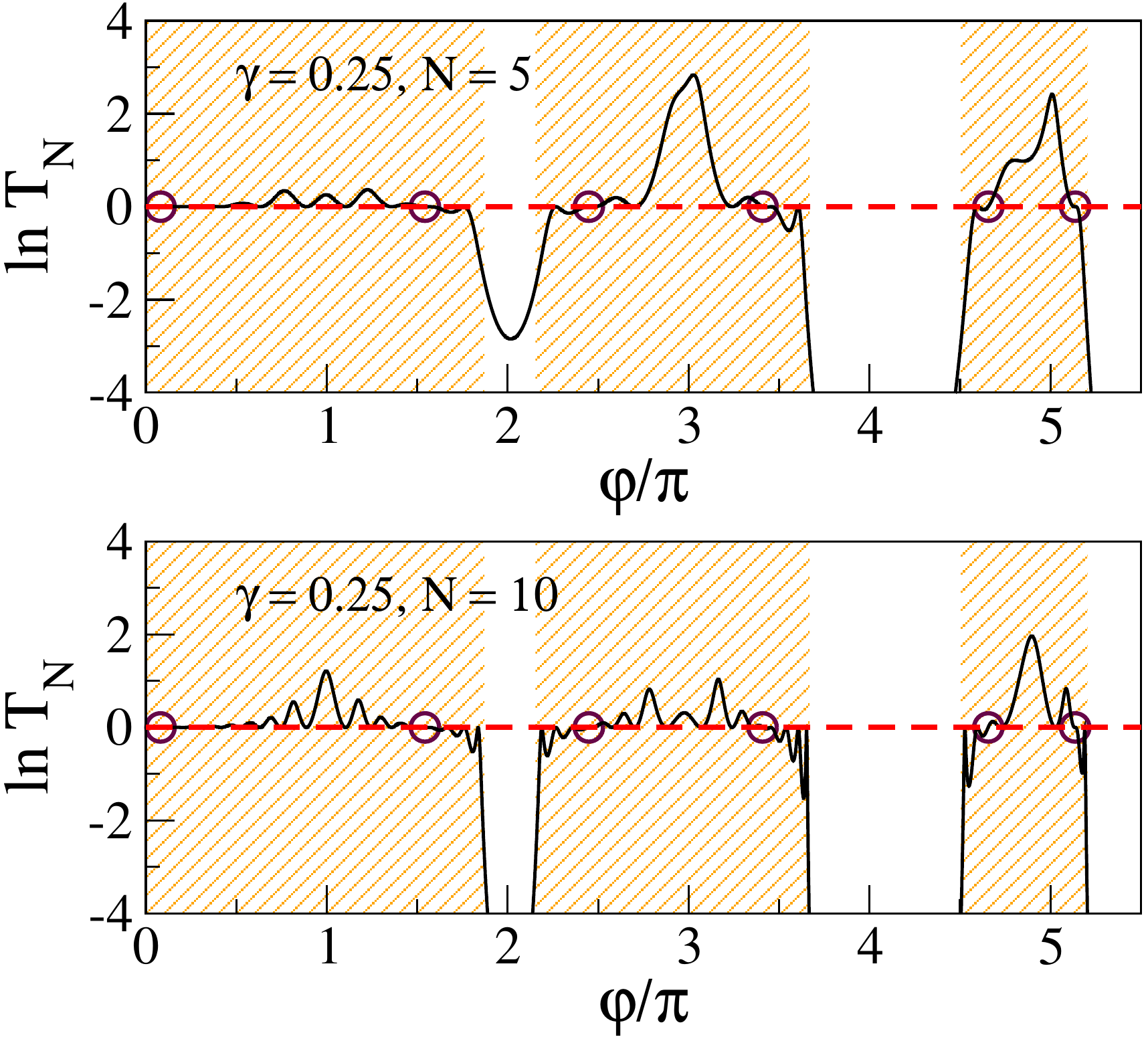}
\caption{The logarithm of $T_N$ defined by Eq.~\eqref{eq:PTMQS-TN} as a function of $\varphi$ for $\gamma=0.25$ and different values $N$ of unit cells. Shaded regions correspond to spectral bands, and circles stand for the borders between the regions with $T_N\leqslant1$ and $T_N\geqslant1$.}\label{fig:LnTN-Phi}
\end{figure}

%%%%%%%%%%%%%%%%%%%%%%%%%%%%%%%%%%%%%%%%%%%%%%%%%%%%%%%%%%%%%%%%%%%%%
\textbf{Transmittance -}
%%%%%%%%%%%%%%%%%%%%%%%%%%%%%%%%%%%%%%%%%%%%%%%%%%%%%%%%%%%%%%%%%%%%%
With the standard transfer matrix approach we derived analytical expression for the transmittance $T_N$ of our model consisting of $N$ unit $(a,b)$ cells and connected to perfect $c\,$-leads by the impedance $Z$,
\begin{eqnarray}
&&T_N=\bigg[1-\frac{\gamma^2\sin^2(N\mu_B)}{4\left(1+\gamma^2\right)^{2}\sin^2\mu_B}\,
\mathcal{F}(\gamma,\varphi)\mathcal{F}(-\gamma,\varphi)\bigg]^{-1},\quad\label{eq:PTMQS-TN}\\
&&\mathcal{F}(\gamma,\varphi)=\gamma^2\sinh(\gamma\varphi)-\gamma\sin\varphi+2(\cos\varphi-\mathrm{e}^{-\gamma\varphi}).\label{eq:PTMQS-Fdef}
\end{eqnarray}
Here $\mathcal{F}(-\gamma,\varphi)<0$ for any value of $\varphi$, unlike $\mathcal{F}(\gamma,\varphi)$ that can be either positive or negative as a function of $\varphi$. Two examples of dependence $T_N(\varphi)$ are given in Fig.~\ref{fig:LnTN-Phi}.
%
%%%%%%%%%%%%%%%%%%%%%%%%%%%%%%%%%%%%%%%%%%%%%%%%%%%%%%%%%%%%%%%%%%%%%
%%%%%%%%%%%%%%%%%%%%%%%%%%%%%%%%%%%%%%%%%%%%%%%%%%%%%%%%%%%%%%%%%%%%%

\begin{figure}[t]
\centering
\includegraphics[scale=0.25]{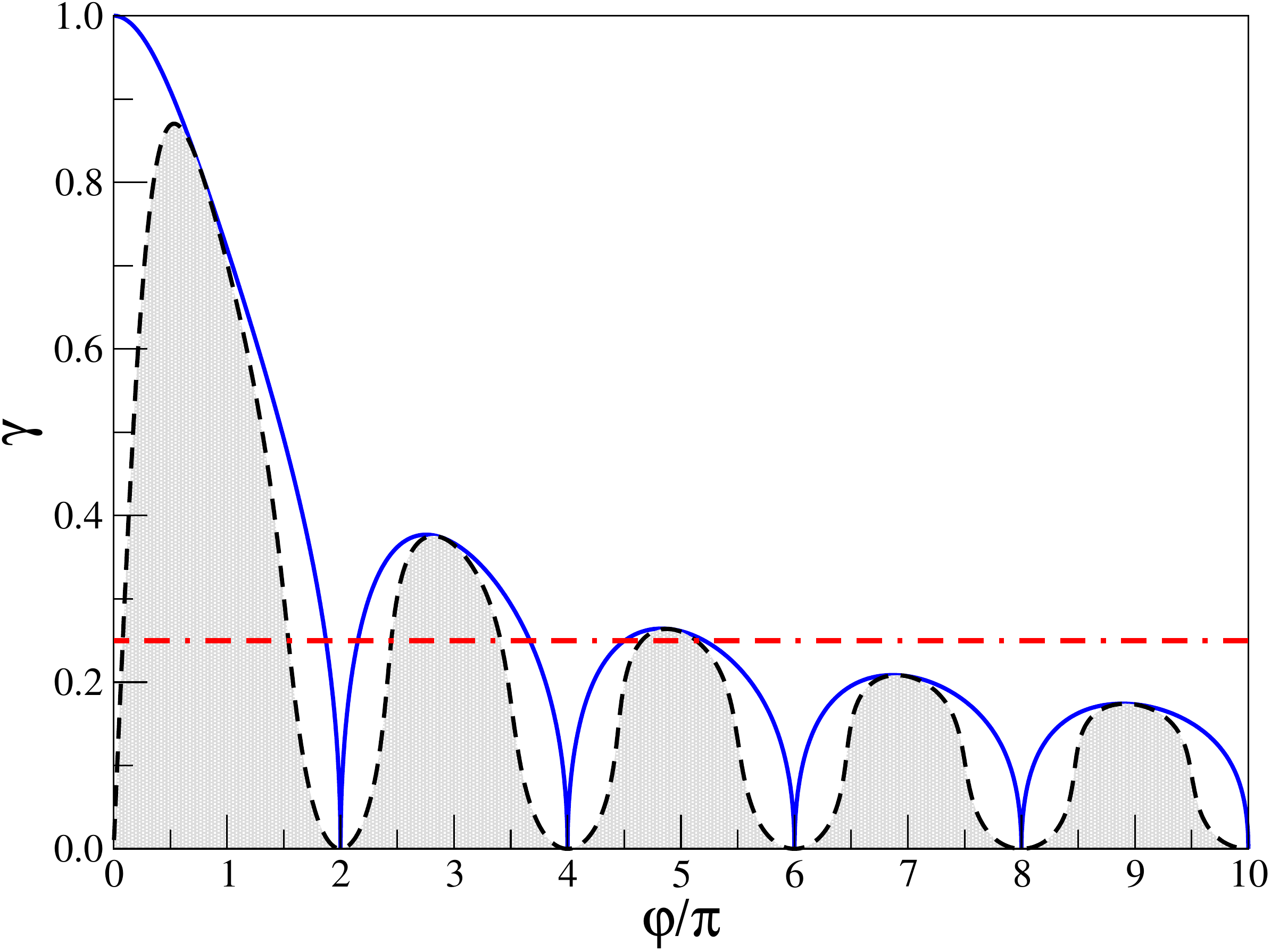}
\caption{Spectral bands and gaps in dependence on the loss/gain parameter $\gamma$. Full curve shows the border $\mu_B=0$ above which $\mu_B$ is imaginary (gaps). Below this curve $\mu_B$ is real. Between full and dashed curve $\mu_B$ is real and $T_N\leqslant1$. Inside dashed regions $\mu_B$ is real and $T_N\geqslant1$. The dependence $\varphi=\varphi_s(\gamma)$ is depicted by the dashed curve on which $T_N=1$. The horizontal dash-dotted line is shown for $\gamma=0.25$.}\label{fig:PTMQS-RegionSuperTrans}
\end{figure}

The analysis shows that a distinctive property of the transmittance is that every spectral band consists of the central region with $T_N\geqslant1$ and those with $T_N\leqslant1$, see Figs.~\ref{fig:LnTN-Phi} and \ref{fig:PTMQS-RegionSuperTrans}. At the borders between these regions (circles in Fig.~\ref{fig:LnTN-Phi}) the transmission is perfect, $T_N =1$, and this happens due to vanishing the function $\mathcal{F}(\gamma,\varphi)$. The corresponding values $\varphi_s(\gamma)$ for which $\mathcal{F}(\gamma,\varphi_s)=0$ can be compared with {\it exceptional points} emerging in the $\mathcal{P}\mathcal{T}$-symmetric models (see, e.g., Ref.~\cite{Yo12} and references therein). However, in our case the Bloch index $\mu_B(\gamma,\varphi_s)$ does not vanish, in contrast to truly exceptional points for which {\it both} conditions, $T_N=1$ and $\mu_B=0$, are fulfilled. These conditions are known as that providing \emph{invisibility} of a model in scattering process. The only situation for the invisibility to occur in our setup is when two conditions, $\mathcal{F}(\gamma,\varphi)=0$ and $\mu_B(\gamma,\varphi)=0$, meet. As one can see in Fig.~\ref{fig:PTMQS-RegionSuperTrans}, the sequence of exceptional points $\varphi=\varphi_{sp}\equiv\varphi_s(\gamma_{sp})$ emerges for the corresponding sequence of $\gamma=\gamma_{sp}$ for which two curves $\varphi=\varphi_s(\gamma)$ and $\mu_B(\gamma,\varphi)=0$ touch each other. Note that at every fixed value of $\gamma=\gamma_{sp}$ the corresponding exceptional point $\varphi_{sp}$ is located on the top of the highest band.
%
%%%%%%%%%%%%%%%%%%%%%%%%%%%%%%%%%%%%%%%%%%%%%%%%%%%%%%%%%%%%%%%%%%%%%
%%%%%%%%%%%%%%%%%%%%%%%%%%%%%%%%%%%%%%%%%%%%%%%%%%%%%%%%%%%%%%%%%%%%%

Apart from the specific points $\varphi_s(\gamma)$, the perfect transmission, $T_N=1$, occurs due to \emph{Fabry-Perot resonances} located within spectral bands. They are defined by the condition $\sin(N\mu_B)/\sin\mu_B=0$ specifying the resonant values $\mu_B^{res}=m\pi/N$ ($m=1,2,3,\ldots,N-1$) of the Bloch index $\mu_B$. Since the spectrum $\mu_B(\omega)$ is restricted by $\mu_B^{(max)}$, the total number of such resonances in a given spectral band can vary from zero to $2(N-1)$, depending on the band number and loss/gain parameter $\gamma$. Specifically, for a fixed $\gamma$ the higher spectral band contains the smaller number of the Fabry-Perot resonances. This result is principally different from what happens in an array of bi-layers with real and positive optical parameters, where $N-1$ number of the Fabry-Perot resonances emerge in any spectral band.

%%%%%%%%%%%%%%%%%%%%%%%%%%%%%%%%%%%%%%%%%%%%%%%%%%%%%%%%%%%%%%%%%%%%%
\textbf{Unidirectional reflectivity -}
%%%%%%%%%%%%%%%%%%%%%%%%%%%%%%%%%%%%%%%%%%%%%%%%%%%%%%%%%%%%%%%%%%%%%
The expressions for the left and right reflectances, $R_N^{(L)}$ and $R_N^{(R)}$, read
\begin{eqnarray}
&&\frac{R_N^{(L)}}{T_N}=\frac{\gamma^2\sin^2(N\mu_B)}{4\left(1+\gamma^2\right)^{2}\sin^2\mu_B}\,
\mathcal{F}^2(\gamma,\varphi)\,;\label{eq:PTMQS-lhR}\\
&&\frac{R_N^{(R)}}{T_N}=\frac{\gamma^2\sin^2(N\mu_B)}{4\left(1+\gamma^2\right)^{2}\sin^2\mu_B}\,
\mathcal{F}^2(-\gamma,\varphi)\,.\label{eq:PTMQS-rhR}
\end{eqnarray}
Note that $T_N(-\gamma)=T_N(\gamma)$ and $R_N^{(L)}(-\gamma)=R_N^{(R)}(\gamma)$. It is remarkable that $T_N(\gamma)$ and $R_N^{(L)}(\gamma), R_N^{(R)}(\gamma)$ satisfy the famous relation,
\begin{equation}
\left|1-T_N\right|=\sqrt{R_N^{(L)}R_N^{(R)}},
\end{equation}
that is known to occur in $\mathcal{P}\mathcal{T}$-symmetric models (see, e.g., Ref.~\cite{GCS12}).

\begin{figure}[t]
\centering
\includegraphics[scale=0.45]{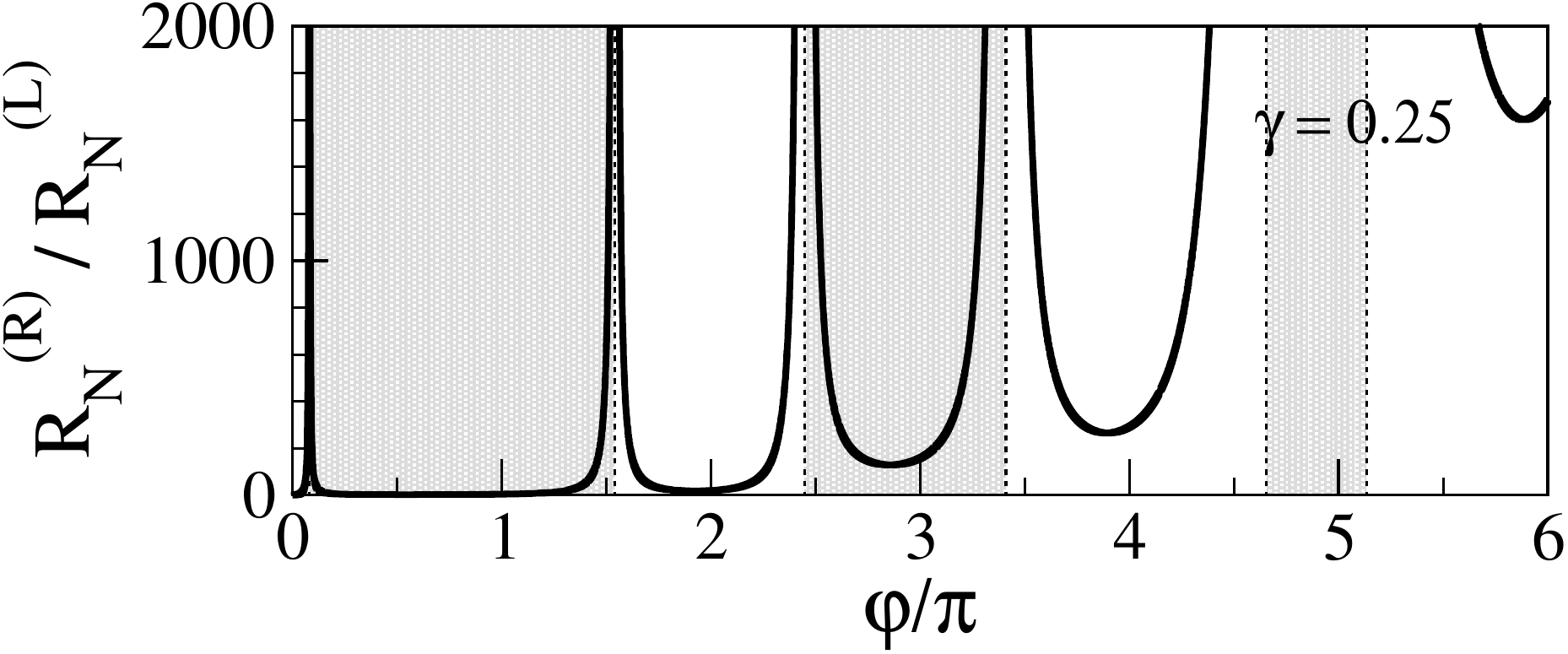}
\caption{Ratio right-to-left reflectances (full curve) versus $\varphi$ for $\gamma=0.25$. Dashed regions are those where $\mu_B$ is real and $T_N>1$, see  Fig.~\ref{fig:PTMQS-RegionSuperTrans}.}\label{fig:Fig5}
\end{figure}

Since at the specific points $\varphi=\varphi_s(\gamma)$ the function $\mathcal{F}(\gamma,\varphi)$ vanishes, the left reflectance \eqref{eq:PTMQS-lhR} also vanishes, however, the right one \eqref{eq:PTMQS-rhR} remains finite. For $\gamma\ll1$ and $\gamma\exp(\gamma\varphi)\gg1$ it can be estimated as follows,
\begin{equation}
R_N^{(R)}\approx16\,\gamma^{-4}\sin^2\left(N\mu_B\right).
\end{equation}
This effect is known as the \emph{unidirectional reflectivity} \cite{Lo11}. It is one of the most important properties of scattering in the $\mathcal{P}\mathcal{T}$-symmetric systems. The ratio between right and left reflectances,
\begin{equation}\label{eq:PTMQS-Rratio}
R_N^{(R)}/R_N^{(L)}=\mathcal{F}^2(-\gamma,\varphi)/\mathcal{F}^2(\gamma,\varphi),
\end{equation}
is shown in Fig.~\ref{fig:Fig5} in dependence on $\varphi$ for $\gamma=0.25$. Remarkably, this ratio is $N$-independent, whereas at the \emph{Fabry-Perot resonances} where $\sin(N\mu_B)/\sin\mu_B=0$, both reflectances expectedly vanish, $R_N^{(L)}=R_N^{(R)}=0$.
%
%%%%%%%%%%%%%%%%%%%%%%%%%%%%%%%%%%%%%%%%%%%%%%%%%%%%%%%%%%%%%%555
%%%%%%%%%%%%%%%%%%%%%%%%%%%%%%%%%%%%%%%%%%%%%%%%%%%%%%%%%%%%%%%%%%%%%

As to the {\it exceptional points} $\varphi=\varphi_{sp}$ ($\gamma=\gamma_{sp}$) where our setup can be treated as invisible, the left reflectance vanishes, $R_N^{(L)}=0$, while the right one is very large,
\begin{equation}\label{eq:PTMQS-rhR-Spoints}
R_N^{(R)}=\frac{4\gamma^2 N^2}{\left(1+\gamma^2\right)^{2}}\left[\cosh(\gamma\varphi)-\cos\varphi\right]^2
\approx\gamma^2N^2\mathrm{e}^{2\gamma\varphi}.
\end{equation}
The estimate is given for $\gamma_{sp}\ll1$ and, hence, $\gamma_{sp}\varphi_{sp}\gg1$.

Within the \emph{spectral gaps} where the Bloch index is purely imaginary ($\mu_B=i|\mu_B|$), for a sufficiently long structure, $N|\mu_B|\gg1$, the transmittance \eqref{eq:PTMQS-TN} is exponentially small. In such a situation the left reflectance \eqref{eq:PTMQS-lhR} is larger than $T_N$, however smaller than $1$. The right reflectance \eqref{eq:PTMQS-rhR} can be shown to exceed unity. Summarizing, one can write, $T_N<R_N^{(L)}<1<R_N^{(R)}$. Both reflectances increase and eventually saturate inside the gaps for $N\to\infty$. However, both $R_N^{(L)}\ll1$ and $R_N^{(R)}\gg1$ remain to be valid. Thus, the ratio $R_N^{(L)}/R_N^{(R)}$ can reach extremely small values of the order of $10^{-7}$.

%%%%%%%%%%%%%%%%%%%%%%%%%%%%%%%%%%%%%%%%%%%%%%%%%%%%%%%%%%%%%%%%%%%%%
\textbf{Conclusions -}
%%%%%%%%%%%%%%%%%%%%%%%%%%%%%%%%%%%%%%%%%%%%%%%%%%%%%%%%%%%%%%%%%%%%%
We have analytically studied the model that reveals the $\mathcal{P}\mathcal{T}$-symmetric transport although its Hamiltonian, in general, is not the $\mathcal{P}\mathcal{T}$-symmetric one. Our analysis exhibits that such a situation emerges in the bi-layer optical setup which without loss/gain is known as the \emph{quarter stack}, provided all layers are {\it perfectly matched} (no reflections due to boundary conditions).

In the absence of loss/gain terms, all spectral bands are touched thus creating a perfect transmission for any wave frequency. When the balanced loss and gain are alternatingly included in all layers, the spectral gaps emerge between the bands, and the total number of bands is defined by the loss/gain parameter $\gamma$ only.

Inside the spectral bands the Bloch index $\mu_B$ appears to be real as it happens in the known models with the $\mathcal{P}\mathcal{T}$-symmetric Hamiltonians. The analysis shows that each of the spectral bands consists of central region where the transmission coefficient is greater than one, $T_N\geqslant1$,  and two side regions with $T_N\leqslant1$. At the borders between these regions the transmission is perfect, $T_N=1$, however, $\mu_B$ does not typically vanish as happens in the known $\mathcal{P}\mathcal{T}$-symmetric models. These borders are of specific interest since one of the reflectances vanishes, thus leading to the so-called "unidirectional reflectivity", the effect which is important both from theoretical and experimental points of view. The ratio between left and right reflectances is specified by the loss/gain parameter $\gamma$ and the wave frequency $\omega$, being independent of the number $N$ of cell units. It is important that a strong difference between left and right reflectances occurs in a relatively large wave-propagation region where the Bloch index $\mu_B$ is real.

The analytical expressions display that inside the spectral bands the \emph{Fabry-Perot resonances} emerge in spite of the presence of loss and gain. These resonances exist both in the regions with $T_N>1$ and $T_N<1$. It is also worthwhile that the invisibility (when both relations, $T_N=1$ and $\mu_B=0$ hold) is realized only for a quite specific values $\gamma_{sp}$ and for the corresponding wave frequencies located in the highest spectral bands.

In conclusion, our study unexpectedly manifests that the $\mathcal{P}\mathcal{T}$-symmetric properties may occur in non-$\mathcal{P}\mathcal{T}$-symmetric systems, the fact that may be important in view of experimental realizations of matched quarter stacks, as well as for the theory of the $\mathcal{P}\mathcal{T}$-symmetric transport.

%%%%%%%%%%%%%%%%%%%%%%%%%%%%%%%%%%%%%%%%%%%%%%%%%%%%%%%%%%%%%%%%%%%%%
\textbf{Acknowledgments -}
%%%%%%%%%%%%%%%%%%%%%%%%%%%%%%%%%%%%%%%%%%%%%%%%%%%%%%%%%%%%%%%%%%%%%
The authors are thankful to L.~Deych, T.~Kottos and A.~Lisyansky for fruitful discussions. This work was supported by the CONACYT (M\'exico) grant No.~CB-2011-01-166382 and by the VIEP-BUAP grant IZF-EXC15-G. D.N.C. acknowledges the support by NSF (Grant No. ECCS-1128520), and AFOSR (Grant No. FA9550-14-1-0037).
%%%%%%%%%%%%%%%%%%%%%%%%%%%%%%%%%%%%%%%%%%%%%%%%%%%%%%%%%%%%%%%%%%%%%

%%%%%%%%%%%%%%%%%%%%%%%%%%%%%%%%%%%%%%%%%%%%%%%%%%%%%%%%%%%%%%%%%%%%%

%%%%%%%%%%%%%%%%%%%%%%%%%%%%%%%%%%%%%%%%%%%%%%%%%%%%%%%%%%%%%%%%%%%%%

\end{document}